\documentclass[12pt]{article}

\usepackage{amssymb}
\usepackage{amsmath}
\usepackage[hiresbb]{graphicx}
\usepackage{amsthm}   
\usepackage{authblk}   
\usepackage{lscape}   
\usepackage{url}
\usepackage{multirow}
\usepackage{here}
\usepackage{xcolor}
\usepackage{comment}
\usepackage{setspace}
\usepackage{natbib}
\makeatletter

\@addtoreset{equation}{section}
\makeatother

\makeatletter
\def\mojiparline#1{
    \newcounter{mpl}
    \setcounter{mpl}{#1}
    \@tempdima=\linewidth
    \advance\@tempdima by-\value{mpl}zw
    \addtocounter{mpl}{-1}
    \divide\@tempdima by \value{mpl}
    \advance\kanjiskip by\@tempdima
    \advance\parindent by\@tempdima
}
\makeatother
\def\linesparpage#1{
    \baselineskip=\textheight
    \divide\baselineskip by #1
}

\allowdisplaybreaks[4]

\newcommand{\bld}{\boldsymbol}

\usepackage[top=1in,bottom=1in,left=1in,right=1in]{geometry}

\title{Nonparametric Bayesian Adjustment of Unmeasured Confounders in Cox Proportional Hazards Models
}
\author[1]{Shunichiro Orihara \thanks{address: 6-1-1 Shinjuku, Shinjuku-ku, Tokyo 160-8402, Japan; email: orihara@tokyo-med.ac.jp}}
\author[2]{Shonosuke Sugasawa}
\author[3]{Tomohiro Ohigashi}
\author[4]{Tomoyuki Nakagawa}
\author[1]{Masataka Taguri}

\affil[1]{Department of Health Data Science, Tokyo Medical University, Tokyo, Japan}
\affil[2]{Graduate School of Economics, Keio University, Tokyo, Japan}
\affil[3]{Department of Information and Computer Technology, Faculty of Engineering, Tokyo University of Science, Tokyo, Japan}
\affil[4]{School of Data Science, Meisei University, Tokyo, Japan}

\date{}

\begin{document}
\linesparpage{25}
\allowdisplaybreaks[4]
\begin{singlespace}
\maketitle
\end{singlespace}
\section*{Abstract}
In observational studies, unmeasured confounders present a crucial challenge in accurately estimating desired causal effects. To calculate the hazard ratio (HR) in Cox proportional hazard models for time-to-event outcomes, two-stage residual inclusion and limited information maximum likelihood are typically employed. However, these methods are known to entail difficulty in terms of potential bias of HR estimates and parameter identification. This study introduces a novel nonparametric Bayesian method designed to estimate an unbiased HR, addressing concerns that previous research methods have had. Our proposed method consists of two phases: 1) detecting clusters based on the likelihood of the exposure and outcome variables, and 2) estimating the hazard ratio within each cluster. Although it is implicitly assumed that unmeasured confounders affect outcomes through cluster effects, our algorithm is well-suited for such data structures. The proposed Bayesian estimator has good performance compared with some competitors.

\vspace{0.5cm}
\noindent
{\bf Keywords}: general Bayes, invalid instrumental variable, Mendelian randomization, UK Biobank, weak instrumental variable

\section{Introduction}
In observational studies, an unmeasured confounders problem, where certain confounders are not observed, results in biased causal effect estimates. In such cases, the instrumental variable (IV) method is one solution to overcome this issue. In the fields of biometrics and related areas, several theoretical results and applications of IV methods have emerged in recent years. Mendelian Randomization (MR), an IV method that typically uses single nucleotide polymorphisms (SNPs) as IVs, is one well-known example.

Time-to-event outcomes are frequently encountered in biometrics and related fields, including MR contexts \citep{Je2021,Wa2018}. The Cox Proportional Hazards Model \citep{Co1972} is typically used to analyze these outcomes by summarizing the causal effect as the hazard ratio, which represents the difference in the occurrence speed of the interested event. IV methods applicable to the CPHM have recently been proposed from various perspectives. \cite{Ki2021} and \cite{Cu2023} propose the weighted estimating equation-based estimator, which can be interpreted as the causal hazard ratio. This approach has the attractive feature that the estimated hazard ratio can be viewed as a causal effect, however, it can only be applied when the IV is a single and binary variable. Since many and continuous IVs are typically considered in MR, applying their methods is not straightforward. Expanding the Two-Stage Residual Inclusion (2SRI) approach, an IV method to applicable to binary outcomes \citep{Te2008}, \cite{Ma2019} proposed a method based on the frailty model \citep{Ni1992}, which considers unobserved variability as frailty. While they explored useful properties of their estimator under continuous exposure conditions, 2SRI may still produce biased and unstable causal effects in certain scenarios \citep{Ba2017,Wa2018,Or2023}, especially in weak IV situations commonly encountered in MR \citep{Or2024a}.

\cite{Ba2017} and \cite{Or2023} demonstrated that a full-likelihood approach yields more accurate estimates than 2SRI. The Limited Information Maximum Likelihood (LIML) estimator, another IV method applicable to binary outcomes \citep{Wo2014}, employs a full-likelihood approach. It is well-known that LIML produces robust estimates in weak IV situations \citep{Bu2017,Gr2017,Or2022} proposed a LIML-based hazard ratio estimator that can be applied to various types of instruments and exposure variables. However, their proposed method requires strong assumptions: 1) a common normal distribution for the unmeasured confounders, and 2) the specification of a variance parameter for the unmeasured confounders. 

In this study, we introduce a novel estimating procedure for the hazard ratio, which is derived from a perspective distinct from the aforementioned methods. The LIML-based method partly shares the same foundational approach as our proposed method, since both use likelihood; however, our method is grounded in the Bayesian paradigm. Unlike the LIML approach, our proposed method does not suffer from the two limitations associated with the distribution of unmeasured confounders. Specifically, our method employs nonparametric Bayesian techniques with cluster construction to account for unmeasured confounders; that is, the distributions of unmeasured confounders may vary among subjects. To appropriately estimate the hazard ratio, our approach involves two phases: 1) detecting clusters based on the likelihood of the exposure and outcome variables, and 2) estimating the hazard ratio within each identified cluster. To properly implement the second step within the Bayesian paradigm, we employ general Bayes methods \citep{Bi2016} that eliminate the need for baseline hazard modeling. While Bayesian methods are used in related topics, our unique contribution is the application of general Bayes techniques to the CPHM.

The remainder of this manuscript is organized as follows: In Section 2, we introduce the models and assumptions used in this study. Specifically, we discuss scenarios where unmeasured confounders exhibit cluster construction, which is a key insight of this manuscript. In Section 3, we detail the proposed Bayesian methods and their sampling algorithms, including an in-depth examination of the procedures. In Section 4, we assess the properties of the proposed method through simulation experiments.


\section{Preliminaries}
\subsection{Cox proportional hazards models with clustering structures}
For $i=1,\ldots,n$, let $T_i$ be a survival time, $\delta_i$ is a censoring indicator whether an event is observed ($\delta_{i}=1$) or not ($\delta_{i}=0$), and $A_i$ be an exposure variable.
As auxiliary information, let $\bld{x}_i$ be vectors of covariates. Further, let $\lambda_i(t)$ be a hazard function of the $i$th subject. 
We then consider a continuous model for exposure and the Cox proportional hazards model for outcome, described as
\begin{align}
A_{i}&=\alpha_{i}+\bld{x}^{\top}_{i}\bld{\alpha}_{xi}+\varepsilon_{i}, \label{for1_1}\\
\lambda_{i}(t)&=\lambda_{0i}(t)\exp\left(\tilde{\bld{x}}_{i}^{\top}\bld{\beta}\right), \label{for1_2}
\end{align}
where $\varepsilon_i\sim N(0, \sigma_i^2)$ is an error term, and $\lambda_{0i}(t)$ is subject-specific effect and baseline hazard function. Also, $\tilde{\bld{x}}$ represents the vector of explanatory variables for the outcome model. For instance, $\tilde{\bld{x}}$ includes $a$ and $\bld{x}$; $\tilde{\bld{x}}^{\top}=\left(a,\bld{x}^{\top}\right)$ and $\bld{\beta}^{\top}=\left(\beta_{a},\bld{\beta}_{x}^{\top}\right)$. In this setting, we are interested in the treatment effect $\beta_{a}$. Note that more complex model constructions can be considered for both models; for instance, the interaction term between the exposure and confounders can be included in model (\ref{for1_2}).

The model (\ref{for1_1}) and (\ref{for1_2}) have subject-specific parameters, $\bld{\alpha}_{xi}$, $\sigma_i^2$ and $\lambda_{0i}(t)$, which cannot be identified without any structural assumptions. 
Here, we introduce nonparametric Bayesian approach to handle the high-dimensionality of these parameters. 
Specifically, we assume that
\begin{equation}\label{eq:DP}
(\bld{\alpha}_{xi},\sigma_i^2,\lambda_{0i})\mid P \sim P, \ \ \ \ \ \ 
P\sim {\rm DP}(G_0, \gamma),   \ \ \ \ \ \  i=1,\ldots,n,
\end{equation}
where $P$ is a discrete random probability measure and ${\rm DP}(G_0,\gamma)$ is the Dirichlet process prior \citep{Es1998,Co2008,Mu2015} for $P$ with precision parameter $\gamma$ and base measure $G_0$.
When there are cluster structures within the data, the Dirichlet process prior can estimate parameters with clustering, without requiring any information about the clusters or their sizes. The size of each cluster is controlled by $\gamma$, known as the `precision parameter'.

The probability measure $P$ has the expression 
$$
P(\cdot) = \sum_{k=1}^{\infty} w_k \delta_{m_k}(\cdot),
$$
where $w_k= v_k\prod_{\ell=1}^{k-1}(1-v_{\ell})$, $v_k\sim {\rm Be}(1, \gamma)$, $m_k\sim G_0$, and $\delta_a(\cdot)$ denotes the Dirac measure on $a$.
Suppose we have $n$ items $\{1,2,\ldots,n\}$ independently sampled from ${\rm DP}(G_0,\gamma)$ and partitioned into $K_n$ distinct groups $(C_1,\ldots,C_{K_n})$, where $N_k$ represents the cardinality of the $k$th cluster for $k=1,\ldots,K_n$.
Let $s_1,\ldots,s_n \in \{1,\ldots,K_n\}$ be latent assignment of each subject.
Then, the joint probability of $S=(s_1,\ldots,s_n)$ is given by 
\begin{equation}\label{eq:DP1}
P(s_1,\ldots,s_n; \gamma)=\prod_{i=1}^n p(s_i\mid s_1,\ldots,s_{i-1}; \gamma),
\end{equation}
where the conditional probability exhibits the well-known Chinese restaurant process:  
\begin{equation}\label{eq:DP2}
\begin{split}
&p(s_i=k\mid s_1,\ldots,s_{i-1};\gamma) = \frac{N_k(i-1)}{i-1+\gamma}
, \ \ \ k=1,\ldots,K_{i-1},\\
&p(s_i=K_{i-1}+1\mid s_1,\ldots,s_{i-1};\gamma)=
\frac{\gamma}{i-1+\gamma},
\end{split}
\end{equation}
where $N_k(i-1)=\sum_{j=1}^{i-1}{\rm I}(s_j=k)$ is the size of the $k$th cluster induced by $s_1,\ldots,s_{i-1}$, and  $I(\cdot)$ is an indicator function.

\subsection{Confounding Bias as Cluster Effect}
As described in the previous section, the use of the Dirichlet process introduces the latent partition structures for $n$ subjects. 
For the $k$th cluster, the model (\ref{for1_1}) and (\ref{for1_2}) can be expressed as 
$$
A_i\sim N(\alpha_{0k}+\bld{x}_i^{\top}\bld{\alpha}_{xk},\sigma_k^2),  \ \ \ \ \ \ 
\lambda_i(t)=\lambda_{0k}(t)\exp\left(a_i\beta_{a}+\bld{x}_i^{\top}\bld{\beta}_{x}\right).
$$
Hence, each cluster has different intercept $\alpha_{0k}$, coefficients $\bld{\alpha}_{xk}$, variance $\sigma_k^2$ and baseline hazard $\lambda_{0k}(t)$, which can be interpreted as a cluster effect resulting from unmeasured confounders.

\begin{figure}[h]
\begin{center}
\begin{tabular}{c}
\includegraphics[width=14cm]{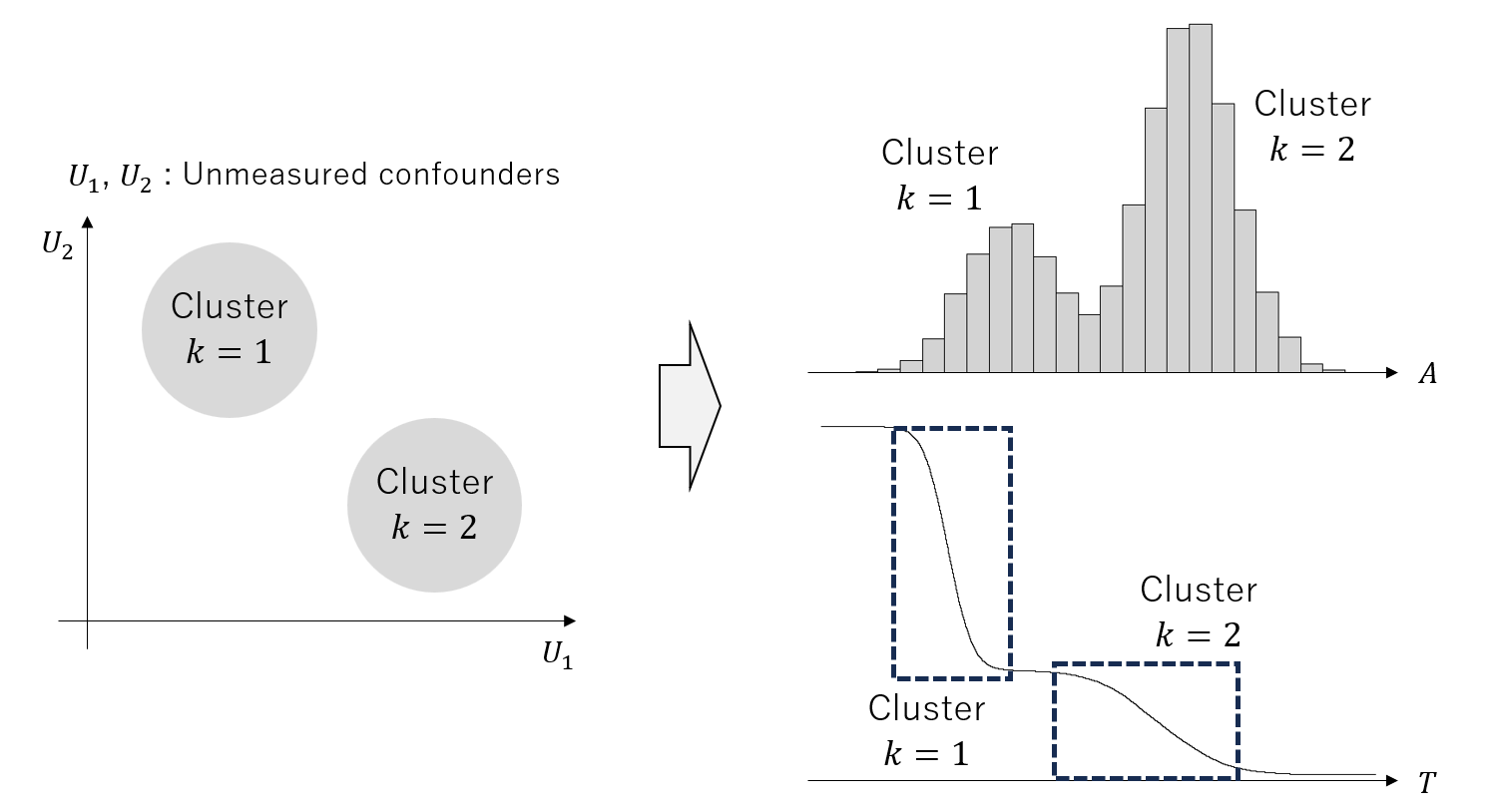}
\end{tabular}
\caption{Image of the impact of cluster effects as confounding bias}
\label{fig1}
\end{center}
\end{figure}

For instance, consider a scenario where there are only two clusters: $k\in\{1,2\}$ (see Figure \ref{fig1}). In this example, the subject who has a larger value of $\alpha_{k}$ tends to have a higher value of $\lambda_{0k}(t)$ in the later time period. In such cases, subjects with smaller exposure values tend to experience the event of interest more early. This suggests that the treatment effect $\beta_{a}$ could be biased if the cluster effects are not considered. This bias is a well-known phenomenon referred to as `confounding bias.'

In this context, adjusting for (unmeasured) confounder effects is equivalent to conducting analyses considering the cluster effect. In other words, to accurately estimate the treatment effect $\beta_{a}$, it is essential to account for the cluster effect. This setting assumes that unmeasured confounders have discrete supports, essentially the same setting as described by \cite{Mi2018} and \cite{Sh2020}. In this manuscript, using nonparametric Bayesian estimation, we propose a novel method to overcome the unmeasured confounder problem.

\subsection{Necessity of Instrumental Variables}
In the models (\ref{for1_1}) and (\ref{for1_2}), IVs are not explicitly shown. In fact, valid IVs are not necessarily required to construct our proposed estimating procedure; their role differs from that in other IV methods, such as 2SRI. The important consideration is that the likelihoods of both models are plausible.

Here, suppose that the confounder and related coefficients can be decomposed into two parts: $\bld{X}^{\top} = (\bld{Z}^{\top}, \bld{V}^{\top})$ and $\bld{\alpha}_{xk}^{\top} = (\bld{\alpha}_{z}^{\top}, \bld{\alpha}_{vk}^{\top})$. Note that $\bld{Z}$ has the common effect $\bld{\alpha}_{z}$ on the exposure variable across clusters, and this effect is `independent' of the unmeasured confounder effect (i.e., $k$). This characteristic of the variable $\bld{Z}$ relates to the two IV assumptions \citep{Bu2021}:
\begin{description}
\item{\bf IV cond. 1):} IVs are related to the treatment,
\item{\bf IV cond. 2):} IVs are independent of unmeasured confounders.
\end{description}
In other words, $\bld{Z}$ is a common predictor of the exposure variable; it plays an important role in identifying each cluster accurately, while $\bld{V}$ is predictor within each cluster. However, $\bld{Z}$ is not necessarily excluded from the outcome model (\ref{for1_2}) in our proposed method. This represents a violation of the exclusion restriction, which is the last IV condition. Consequently, $\bld{Z}$ may not necessarily be an IV; the acceptance of an invalid IV due to horizontal pleiotropy \citep{Da2018} is permissible.

The model (\ref{for1_2}) can be considered a frailty model \citep{Ni1992} because the baseline hazard functions vary across clusters. In other words, the influence of the cluster effect on the outcome value is mediated solely through the baseline hazard function, not through the treatment effects. This aligns with the homogeneous treatment assumption \citep{Bu2021}. From the model constructions, simple estimating procedure can be constructed.

\section{Nonparametric Bayesian Adjustment}
To estimate the target parameter $\bld{\beta}$ under existence of unmeasured confounders, we propose a novel nonparametric Bayesian estimation method. To construct a Bayesian procedure, it is commonly assumed that data information is reflected in the parameter of interest through a specific likelihood function. However, we assume a partial likelihood for the parameter $\bld{\beta}$. As discussed in \cite{Bi2016}, a loss function that includes the partial likelihood for the CPHM can be justified from a Bayesian perspective. In other words, the partial likelihood can be applied in the same manner as ordinary likelihoods. To proceed with the following discussions, we assume that the censoring time, except for the event of interest, is independent of the event time $t_{i}$ over clusters $k$ given $A$ and $\bld{X}$. This assumption is similar to that made by \cite{Or2022}.

\subsection{Posterior of nonparametric Bayesian adjustment}
Given the cluster assignment, $S$, the outcome model in (\ref{for1_2}) reduces to the following clustered Cox proportional hazards model: 
$$
\lambda_i(t)|(s_i=k)=\lambda_{0k}(t)\exp\left(a_i\beta_{a}+\bld{x}_i^{\top}\bld{\beta}\right),  \ \ \ \ \ k=1,\ldots,K_n.
$$
To make inference on $\bld{\beta}$, we consider the clustered partial likelihood, ${\rm PL}(\bld{\beta}\mid S)=\prod_{k=1}^{K_n}\prod_{i\in C_k} \ell_{ik}(\bld{\beta})$, where 
\begin{align}\label{eq:PL}
\ell_{ik}(\bld{\beta})\equiv \left\{\frac{\exp(a_{i}\beta_{a}+\bld{x}_i^{\top}\bld{\beta}_{x})}{\sum_{\ell \in R_{k}(T_{i})}\exp(a_\ell\beta_{a}+\bld{x}_\ell^{\top}\bld{\beta_{x}})}\right\}^{ \delta_i },
\end{align}
and $R_{k}(t)$ denotes the cluster-wise risk set at time $t$.
A notable feature of the partial likelihood (\ref{eq:PL}) is that it is free from the nuisance baseline hazard function, $\lambda_{0k}(\cdot)$, so that we can consider posterior distribution focusing on $\bld{\beta}$.
Although standard Bayesian inference on censored survival outcome requires a full likelihood (e.g. \cite{Ka2011}) and assuming a prior distribution for the baseline hazard function \citep{Si2003}, a framework called ``general posterior" \citep{Bi2016} enables us to define a valid posterior distribution based on synthetic likelihood like (\ref{eq:PL}). 

On the other hand, the exposure model in (\ref{for1_2}), given the latent partition, can be expressed as 
\begin{equation}\label{eq:model-mix}
\begin{split}
&A_i|(s_i=k)\sim N(\alpha_{0k}+\bld{v}_i^{\top}\bld{\alpha}_{vk}+\bld{z}_i^{\top}\bld{\alpha}_{z}, \sigma_k^2), \ \ \ \ 
k=1,\ldots,K_n.
\end{split}
\end{equation}
This model is similar to, but not identical to, the likelihood assumed in \cite{Or2022}, which assumes that all subjects share the same error distribution ($\sigma_k=\sigma^2$ for all $k$), and that $\sigma$ is a fixed value. 

Let $O_{i} = (T_{i},\delta_{i},A_{i},\bld{x}_i, \bld{z}_i)$  and $O = (O_{1},\dots,O_{n})$ be the set of observations.
Then, the joint (general) posterior can be obtained as 
\begin{equation}\label{eq:pos}
\begin{split}
p(\bld{\beta}, \bld{\alpha}_{0}, \bld{\alpha}_{v},  \bld{\alpha}_z, \Sigma, \gamma, \bld{s} \mid O)
&\propto \pi(\bld{\beta},  \bld{\alpha}_z, \Sigma, \gamma) \pi_{G_0}(\bld{\alpha}_{0}, \bld{\alpha}_{v}, \Sigma) P(s_1,\ldots,s_n;\gamma)\\
&\times \prod_{k=1}^{K_n}\prod_{i\in C_k} \ell_{ik}(\bld{\beta})
\phi\Big(A_i; \alpha_{0k}+\bld{v}_i^{\top}\bld{\alpha}_{vk}+\bld{z}_i^{\top}\bld{\alpha}_{z}, \sigma_k^2\Big),
\end{split}
\end{equation}
where $\pi(\bld{\beta},  \bld{\alpha}_z, \Sigma, \gamma)$ is a prior distribution on (global) parameters and $\pi_{G_0}(\bld{\alpha}_{0}, \bld{\alpha}_{v}, \Sigma)$ is a prior distribution on cluster-wise parameters generated from the baseline distribution $G_0$. 
Here $\phi(x; a, b)$ denotes the density function of the normal distribution with mean $a$ and variance $b$. 
Note that $P(s_1,\ldots,s_n;\gamma)$ and $\ell_{ik}(\bld{\beta})$ are defined in (\ref{eq:DP1}) and (\ref{eq:PL}), respectively. 

\subsection{Posterior computation }
In this section, we assume the prior distributions, $\beta_{a}\sim N(m_{\beta_{a}}, \tau_{\beta_{a}}^2)$, $\bld{\beta}_{x}\sim N(\bld{m}_{\beta_{x}}, \bld{\Sigma}_{\beta_{x}})$,  $\bld{\alpha}_{vk}\sim N(\bld{m}_{\alpha_{vk}}, \bld{\Sigma}_{\alpha_{vk}})$, $\bld{\alpha}_z\sim N(\bld{m}_{\alpha_z}, \bld{\Sigma}_{\alpha_z})$, $\alpha_{0k}\sim N(m_{\alpha_{0}}, \tau_{\alpha_{0}}^2)$, $\sigma_k^2\sim {\rm IG}(a_{\sigma}, b_{\sigma})$ and $\gamma \sim {\rm Ga}(a_{\gamma}, b_{\gamma})$ as default priors. 
We will consider more general priors in the latter section. 
Based on the joint posterior (\ref{eq:pos}), the full conditional distributions of each parameter and latent assignment are obtained as follows:

\begin{itemize}
\item[-](Sampling of $\bld{\alpha}_{z}$)  \ \ 
The full conditional distribution of $\bld{\alpha}_{z}$ is proportional to 
$$
\pi(\bld{\alpha}_{z})\prod_{k=1}^{K_n}\prod_{i\in C_k} 
\phi\Big(A_i; \alpha_{0k}+\bld{v}_i^{\top}\bld{\alpha}_{vk}+\bld{z}_i^{\top}\bld{\alpha}_{z}, \sigma_k^2\Big).
$$

\item[-](Sampling of $\bld{\alpha}_{0}$ and $\bld{\alpha}_{v}$) \ \ For $k=1,\ldots,K_n$, the full conditional distribution of $(\alpha_{0k},\bld{\alpha}_{vk})$ is proportional to 
$$
\pi(\alpha_{0k},\bld{\alpha}_{vk})\prod_{i\in C_k} 
\phi\Big(A_i; \alpha_{0k}+\bld{v}_i^{\top}\bld{\alpha}_{vk}+\bld{z}_i^{\top}\bld{\alpha}_{z}, \sigma_k^2\Big).
$$

\item[-](Sampling of $\Sigma$) \ \ For $k=1,\ldots,K_n$, the full conditional distribution of $\sigma_k^2$ is proportional to 
$$
\pi(\sigma_k^2)\prod_{i\in C_k} 
\phi\Big(A_i; \alpha_{0k}+\bld{v}_i^{\top}\bld{\alpha}_{vk}+\bld{z}_i^{\top}\bld{\alpha}_{z}, \sigma_k^2\Big).
$$

\item[-](Sampling of $S$) \ \ For $i=1,\ldots,n$, the full conditional distribution of $s_i\in \{1,\ldots,K_n\}$, namely, full conditional probability being $s_i=k$ is proportional to 
$$
P(s_1,\ldots,s_n;\gamma)
\ell_{ik}(\beta_{a}, \bld{\beta}_{x})\phi\Big(A_i; \alpha_{0k}+\bld{v}_i^{\top}\bld{\alpha}_{vk}+\bld{z}_i^{\top}\bld{\alpha}_{z}, \sigma_k^2\Big)
$$
for $k=1,\ldots,K_n$.

\item[-](Sampling of $\gamma$) \ \ The full conditional distribution of $\gamma$ is proportional to 
$$
\pi(\gamma) P(s_1,\ldots,s_n;\gamma).
$$
\item[-](Sampling of $\beta_{a}$ and $\bld{\beta}_{x}$) \ \ For $k=1,\ldots,K_n$, the full conditional distribution of $(\beta_{a}, \bld{\beta}_{x})$ is proportional to 
$$
\pi(\beta_{a}, \bld{\beta}_{x})\prod_{i\in C_k} \ell_{ik}(\beta_{a}, \bld{\beta}_{x}),
$$
where $\ell_{ik}(\beta_{a}, \bld{\beta}_{x})$ is given in (\ref{eq:PL}).
\end{itemize}
Note that we use \texttt{Stan} to generate random samples of $\bld{\beta}$ and $\gamma$. Detailed sampling algorithms for sampling of $S$ and $\gamma$ are described in Appendix \ref{appa}.

In the sampling step for $S$ and $\gamma$, the Chinese restaurant process (\ref{eq:DP2}) will be used in the following simulation and real data experiments. Note that according to the above sampling schemes, $(\beta_{a}, \bld{\beta}_{x})$ can differ in each cluster $k$; i.e., $(\beta_{ak}, \bld{\beta}_{xk})$. More details are discussed in Appendix \ref{appb}.

\subsection{Applying Shrinkage Techniques}
As mentioned previously, valid IVs are not necessarily needed to estimate the hazard ratio properly for our proposed method. In other words, all covariates, including $\bld{Z}$, can be included in both models (\ref{for1_1}) and (\ref{for1_2}). However, in the Mendelian Randomization context, over 100 IVs are typically considered. To accurately estimate the hazard ratio, applying shrinkage methods and reducing the number of parameters are beneficial.

In the real data analysis, we apply the horseshoe prior \citep{Ca2010} to more precisely shrink the coefficient estimates near zero to zero when sampling from the posteriors for $\bld{\alpha}_{z}$ and $\bld{\beta}$. Specifically, we consider the following priors for $\pi(\bld{\alpha}_{z})$ and $\pi(\bld{\beta})$ respectively:
\begin{description}
\item[Prior for $\bld{\alpha}_{z}$] For each $\alpha_{z\ell}$,
$$
\alpha_{z\ell}|\psi_{\ell1}\sim N(0,\psi_{\ell1}^2\tau_{1}^2),\ \ \ \psi_{\ell1}\sim C^{+}(0,1),
$$
\item[Prior for $\bld{\beta}$] For each $\beta_{\ell}$,
$$
\beta_{\ell}|\psi_{\ell2}\sim N(0,\psi_{\ell2}^2\tau_{2}^2),\ \ \ \psi_{\ell2}\sim C^{+}(0,1),
$$
\end{description}
where $C^{+}(0,1)$ is Half-Cauchy distribution.

\section{Simulation Experiments}
In this section, we evaluate the performance of our proposed methods compared to several competitors, which will be explained later, using simulation data examples. The number of iterations for all simulation examples was set to 200.
\subsection{Data-generating Mechanism}

First, we will describe the Data-Generating Mechanism (DGM) used in this simulation. We consider one unmeasured confounder, denoted as $U_{i} \stackrel{i.i.d.}{\sim} Multi\left(1, \left(1/2,1/3,1/6\right)\right)$. This unmeasured confounder, labeled as $U=0$, $U=1$, and $U=2$, derives the cluster effect which we cannot identify it directly. Also, we consider one instrumental variable and measured confounder, denoted as $Z_{i1}\stackrel{i.i.d.}{\sim} Gamma(2, 2)$ and $Z_{i2}\stackrel{i.i.d.}{\sim} Ber(0.5)$, respectively. Using these variables, the exposure variable is defined by $A|z_{i1},z_{i2},u_{i}\sim N\left(\alpha_{0}+z_{i1}\alpha_{z1}+z_{i1}\alpha_{uz2},0.5^2\right)$.

In our simulation experiments, we consider two major settings: ``Easy to identify" and ``Hard to identify." In the former setting, the cluster construction can be identified intuitively, whereas it is difficult to identify in the latter. All settings are summarized in Table \ref{tab0}.

The outcome model has somewhat complex DGM. A basic outcome model is
$$
\lambda(t_{i}|a_{i},z_{i2},u_{i})=\lambda_{0u_{i}}\exp\left\{-0.1a_{i}+0.1z_{i2}\right\}.
$$
First, we explain ``Easy to identify" setting. For $U=0$ subjects, $\lambda_{0u_{i}}=0.15$. Additionally, we consider ``latent" event time, denoted as $\tilde{T}_{i}$ is derived from
\begin{align}
\label{late1}
\tilde{\lambda}(t_{i}|a_{i},z_{i2},u_{i}\neq0)=0.005\exp\left\{-0.1a_{i}+0.1z_{i2}\right\}.
\end{align}
If $U=1$ and $\tilde{T}_{i}<65$, $T_{i}=\tilde{T}_{i}$, the outcome model is (\ref{late1}). Else, $\lambda_{0u_{i}}=0.15$. If $U=2$ and $\tilde{T}_{i}<40$, $T_{i}=\tilde{T}_{i}$, the outcome model is (\ref{late1}). Else, $\lambda_{0u_{i}}=0.1$.

Next, we explain ``Hard to identify" setting. For $U=0$ subjects, $\lambda_{0u_{i}}=0.035$. Additionally, we consider ``latent" event time, denoted as $\tilde{T}_{i}$ is derived from
\begin{align}
\label{late2}
\tilde{\lambda}(t_{i}|a_{i},z_{i2},u_{i}\neq0)=0.015\exp\left\{-0.1a_{i}+0.1z_{i2}\right\}.
\end{align}
If $U=1$ and $\tilde{T}_{i}<65$, $T_{i}=\tilde{T}_{i}$, the outcome model is (\ref{late2}). Else, $\lambda_{0u_{i}}=0.1$. If $U=2$ and $\tilde{T}_{i}<40$, $T_{i}=\tilde{T}_{i}$, the outcome model is (\ref{late1}). Else, $\lambda_{0u_{i}}=0.1$.

\begin{figure}[H]
\begin{center}
\begin{tabular}{c}
\includegraphics[width=12cm]{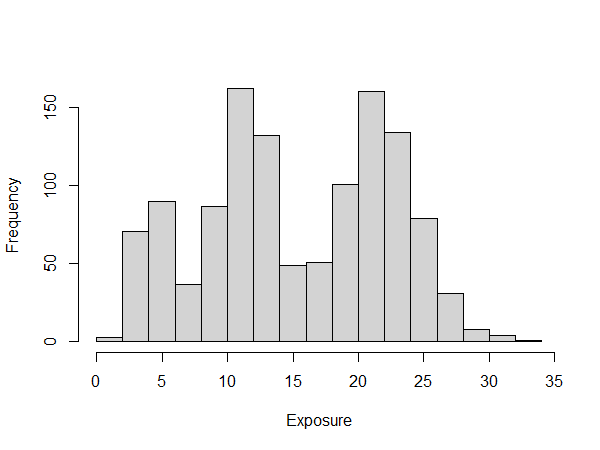}
\end{tabular}
\caption{Histogram of the exposure variable (crude analysis) in a simulation dataset under ``Easy to identify" setting}
\label{fig2}
\end{center}
\end{figure}

\begin{figure}[H]
\begin{center}
\begin{tabular}{c}
\includegraphics[width=12cm]{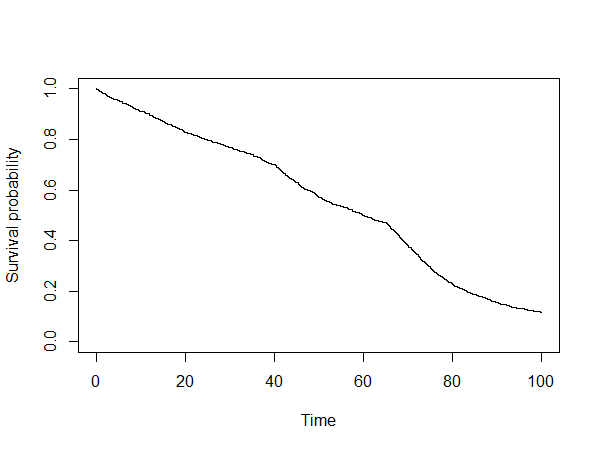}
\end{tabular}
\caption{Kaplan-Meier plot of the outcome variable (crude analysis) in a simulation dataset under ``Easy to identify" setting}
\label{fig3}
\end{center}
\end{figure}

\begin{figure}[H]
\begin{center}
\begin{tabular}{c}
\includegraphics[width=12cm]{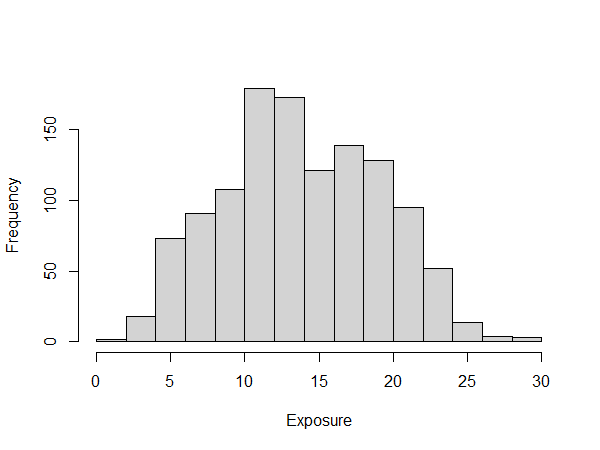}
\end{tabular}
\caption{Histogram of the exposure variable (crude analysis) in a simulation dataset under ``Hard to identify" setting}
\label{fig4}
\end{center}
\end{figure}

\begin{figure}[H]
\begin{center}
\begin{tabular}{c}
\includegraphics[width=12cm]{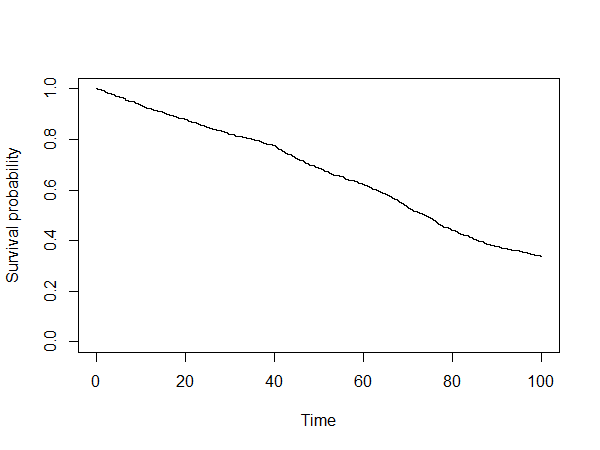}
\end{tabular}
\caption{Kaplan-Meier plot of the outcome variable (crude analysis) in a simulation dataset under ``Hard to identify" setting}
\label{fig5}
\end{center}
\end{figure}

Finally, the censoring is occurred approximately $10$--$15\%$ derived from an exponential distribution. Crude analyses of a simulation dataset is summarized in Figures \ref{fig2}--\ref{fig5}.

\subsection{Estimating Methods}
We considered five methods for estimating the hazard ratio:\ the proposed method and four reference methods.

The proposed method was implemented as described in the preceding sections. In this simulation experiment, the variance parameter is homogeneous (i.e., $\sigma_{k}\equiv\sigma$). Therefore, in the "Sampling of $\Sigma$" step described in Section 3.2, $\sigma$ is sampled from the common posterior for the variance parameter. The iteration time of the MCMC sample is 1200; the first 200 are the burn-in period, and from 201 to 1200 is the sampling step.

Four reference methods were used in this study. The first method, called the ``naive estimator,'' is based on the ordinary CPHM but ignores the unmeasured covariate $U$. This approach leads to biased hazard ratio estimates. The second method, called the ``infeasible estimator,'' includes the unmeasured covariate $U$ in the ordinary CPHM. Although this estimator is consistent and efficient, it cannot be implemented practically; thus, it is referred to as an infeasible estimator. These results are presented only as references. Both the naive and infeasible estimators were implemented using the ``coxph'' function of the survival library in R.

The third method, called ``2SLS," is the ordinary two-stage least squares procedure for estimating the hazard ratio. Specifically, the first stage model is treated as an ordinary linear model. In the second step, predictors of the exposure variables are applied to the ``coxph" function. The standard error of $\beta_{a}$ is directly used to estimate the confidence interval.

The fourth method, called ``2SRI," represents a significant alternative proposed by Mart\'{i}nez-Camblor et al.\ (2019). In simulation experiments, the first stage model is the same as 2SLS. For the second stage model, the ``coxph'' function, with the normal frailty option, is employed. The standard error of $\beta_{a}$ is also directly used to estimate the confidence interval.

\subsection{Performance Metrics}
We evaluated the various methods based on mean, bias, empirical standard error (ESE), root mean squared error (RMSE), coverage probability (CP), and boxplot of estimated parameters from 200 iterations. The bias and RMSE were calculated as follows:
$$
bias=\bar{\hat{\beta}}_{a}-\beta_{a0},\ \ \ RMSE=\sqrt{\frac{1}{200}\sum_{k=1}^{200}\left(\hat{\beta}_{ak}-\beta_{a0}\right)^2},
$$
where $\bar{\hat{\beta}}_{a}=\frac{1}{200}\sum_{k=1}^{200}\hat{\beta}_{ak}$, $\hat{\beta}_{ak}$ is the estimate of each estimator and iteration, and $\beta_{a0}$ ($=-0.1$) is the true value of the log--hazard ratio. The CP refers to the proportion of cases where the confidence interval or credible interval includes $\beta_{a0}$. The boxplot summarize estimated hazard ratio, where the true hazard ratio is $\exp\{\beta_{a0}\}$ ($=\exp\{-0.1\}\approx 0.905$). 

\subsection{Simulation Results}
\subsubsection{Summary results}
All settings are summarized in Tables \ref{tab1}--\ref{tab4}, and Figures \ref{fig6}--\ref{fig9}. Our proposed method produces results relatively similar to the infeasible method. Our method can identify the cluster structure without any prior cluster information and yields highly efficient results. In contrast, the naive method shows an obvious bias, as expected. In strong IV situations, 2SLS is efficient, but it exhibits large variance in weak IV situations; this is a well-known result. 2SRI provides more stable results compared to 2SLS but still shows obvious bias. This bias likely stems from the residuals' inability to accurately capture the distribution of unmeasured confounders. Similar results were discussed in \cite{Or2022}

Our proposed method also demonstrates good CP, which is somewhat conservative but stable, compared to the infeasible method. These results suggest that our proposed methods will work well both in estimation and inference of the hazard ratio with CPHM when there is a clustered unmeasured confounder structure.

\subsubsection{1--shot results}
Additionally summary results, we also consider 1--shot resutlts; especially focused on ``Hard to identify" setting and scenario (a). Results are summarized in Figure \ref{fig10}. Effective sample size of $\beta_{a}$, $\beta_{z1}$, $\beta_{z2}$, $\alpha_{z1}$, and $\sigma^2$ is 443.54, 388.90, 397.54, 116.17, and 383.77, respectively. As expected, sampling from the posterior works well, especially for the coefficients in the outcome model.

Regarding the construction of clusters, nearly exact clustering is achieved, though the number of clusters is quite large (Table \ref{tab5}). This phenomenon is well-known in DP contexts.\cite{Mi2013} In this manuscript, identifying the exact clusters is not the main objective; therefore, this phenomenon is acceptable.


\section{Discussions and Conclusions}
In this study, we introduced a novel nonparametric Bayesian method employing general Bayes techniques to estimate an unbiased hazard ratio in the context of the Cox Proportional Hazards Model (CPHM), particularly when dealing with unmeasured confounders. This method can identify clusters based on the likelihood of the exposure and outcome, and estimate the unbiased hazard ratio estimating within each identified cluster; adjusting unmeasured confounder effects.

Implicit in our approach is the assumption that unmeasured confounders exert cluster effects, a necessary consideration for the suitability of our methods. As discussed in Section 2, this assumption is deemed reasonable under the circumstances. Since we cannot directly identify unmeasured variables, our approach of assuming a cluster-based structure is considered a pragmatic solution. However, future work should focus on relaxing this assumption, exploring more flexible approaches to accommodate a broader range of data structures and confounder characteristics.








\bibliographystyle{chicago}
\bibliography{bibfile}

\begin{thebibliography}{}

\bibitem[\protect\citeauthoryear{Angrist, Imbens, and Rubin}{Angrist
  et~al.}{1996}]{An1996}
Angrist, J.~D., G.~W. Imbens, and D.~B. Rubin (1996).
\newblock Identification of causal effects using instrumental variables.
\newblock {\em Journal of the American statistical Association\/}~{\em
  91\/}(434), 444--455.

\bibitem[\protect\citeauthoryear{Basu, Coe, and Chapman}{Basu
  et~al.}{2017}]{Ba2017}
Basu, A., N.~Coe, and C.~G. Chapman (2017).
\newblock Comparing 2sls vs 2sri for binary outcomes and binary exposures.
\newblock Technical report, National Bureau of Economic Research.

\bibitem[\protect\citeauthoryear{Bissiri, Holmes, and Walker}{Bissiri
  et~al.}{2016}]{Bi2016}
Bissiri, P.~G., C.~C. Holmes, and S.~G. Walker (2016).
\newblock A general framework for updating belief distributions.
\newblock {\em Journal of the Royal Statistical Society Series B: Statistical
  Methodology\/}~{\em 78\/}(5), 1103--1130.

\bibitem[\protect\citeauthoryear{Burgess, Small, and Thompson}{Burgess
  et~al.}{2017}]{Bu2017}
Burgess, S., D.~S. Small, and S.~G. Thompson (2017).
\newblock A review of instrumental variable estimators for mendelian
  randomization.
\newblock {\em Statistical methods in medical research\/}~{\em 26\/}(5),
  2333--2355.

\bibitem[\protect\citeauthoryear{Burgess and Thompson}{Burgess and
  Thompson}{2021}]{Bu2021}
Burgess, S. and S.~G. Thompson (2021).
\newblock {\em Mendelian randomization: methods for causal inference using
  genetic variants}.
\newblock CRC Press.

\bibitem[\protect\citeauthoryear{Carvalho, Polson, and Scott}{Carvalho
  et~al.}{2010}]{Ca2010}
Carvalho, C.~M., N.~G. Polson, and J.~G. Scott (2010).
\newblock The horseshoe estimator for sparse signals.
\newblock {\em Biometrika\/}~{\em 97\/}(2), 465--480.

\bibitem[\protect\citeauthoryear{Conley, Hansen, McCulloch, and Rossi}{Conley
  et~al.}{2008}]{Co2008}
Conley, T.~G., C.~B. Hansen, R.~E. McCulloch, and P.~E. Rossi (2008).
\newblock A semi-parametric bayesian approach to the instrumental variable
  problem.
\newblock {\em Journal of Econometrics\/}~{\em 144\/}(1), 276--305.

\bibitem[\protect\citeauthoryear{Cox}{Cox}{1972}]{Co1972}
Cox, D.~R. (1972).
\newblock Regression models and life-tables.
\newblock {\em Journal of the Royal Statistical Society: Series B
  (Methodological)\/}~{\em 34\/}(2), 187--202.

\bibitem[\protect\citeauthoryear{Cui, Michael, Tanser, and
  Tchetgen~Tchetgen}{Cui et~al.}{2023}]{Cu2023}
Cui, Y., H.~Michael, F.~Tanser, and E.~Tchetgen~Tchetgen (2023).
\newblock Instrumental variable estimation of the marginal structural cox model
  for time-varying treatments.
\newblock {\em Biometrika\/}~{\em 110\/}(1), 101--118.

\bibitem[\protect\citeauthoryear{Dahl, Day, and Tsai}{Dahl
  et~al.}{2017}]{Da2017}
Dahl, D.~B., R.~Day, and J.~W. Tsai (2017).
\newblock Random partition distribution indexed by pairwise information.
\newblock {\em Journal of the American Statistical Association\/}~{\em
  112\/}(518), 721--732.

\bibitem[\protect\citeauthoryear{Davies, Holmes, and Smith}{Davies
  et~al.}{2018}]{Da2018}
Davies, N.~M., M.~V. Holmes, and G.~D. Smith (2018).
\newblock Reading mendelian randomisation studies: a guide, glossary, and
  checklist for clinicians.
\newblock {\em bmj\/}~{\em 362}, k601.

\bibitem[\protect\citeauthoryear{Escobar and West}{Escobar and
  West}{1998}]{Es1998}
Escobar, M.~D. and M.~West (1998).
\newblock Computing nonparametric hierarchical models.
\newblock In {\em Practical nonparametric and semiparametric Bayesian
  statistics}, pp.\  1--22. Springer.

\bibitem[\protect\citeauthoryear{Grover, Del Greco~M, Stein, and
  Ziegler}{Grover et~al.}{2017}]{Gr2017}
Grover, S., F.~Del Greco~M, C.~M. Stein, and A.~Ziegler (2017).
\newblock Mendelian randomization.
\newblock {\em Statistical Human Genetics: Methods and Protocols\/}, 581--628.

\bibitem[\protect\citeauthoryear{Jenkins, Wade, Carslake, Bowden, Sattar, Loos,
  Timpson, Sperrin, and Rutter}{Jenkins et~al.}{2021}]{Je2021}
Jenkins, D.~A., K.~H. Wade, D.~Carslake, J.~Bowden, N.~Sattar, R.~J. Loos,
  N.~J. Timpson, M.~Sperrin, and M.~K. Rutter (2021).
\newblock Estimating the causal effect of bmi on mortality risk in people with
  heart disease, diabetes and cancer using mendelian randomization.
\newblock {\em International Journal of Cardiology\/}~{\em 330}, 214--220.

\bibitem[\protect\citeauthoryear{Kalbfleisch and Prentice}{Kalbfleisch and
  Prentice}{2011}]{Ka2011}
Kalbfleisch, J.~D. and R.~L. Prentice (2011).
\newblock {\em The statistical analysis of failure time data}.
\newblock John Wiley \& Sons.

\bibitem[\protect\citeauthoryear{Kianian, Kim, Fine, and Peng}{Kianian
  et~al.}{2021}]{Ki2021}
Kianian, B., J.~I. Kim, J.~P. Fine, and L.~Peng (2021).
\newblock Causal proportional hazards estimation with a binary instrumental
  variable.
\newblock {\em Statistica Sinica\/}~{\em 31\/}(2), 673.

\bibitem[\protect\citeauthoryear{Mart{\'\i}nez-Camblor, Mackenzie, Staiger,
  Goodney, and O’Malley}{Mart{\'\i}nez-Camblor et~al.}{2019}]{Ma2019}
Mart{\'\i}nez-Camblor, P., T.~Mackenzie, D.~O. Staiger, P.~P. Goodney, and
  A.~J. O’Malley (2019).
\newblock Adjusting for bias introduced by instrumental variable estimation in
  the cox proportional hazards model.
\newblock {\em Biostatistics\/}~{\em 20\/}(1), 80--96.

\bibitem[\protect\citeauthoryear{Miao, Geng, and Tchetgen~Tchetgen}{Miao
  et~al.}{2018}]{Mi2018}
Miao, W., Z.~Geng, and E.~J. Tchetgen~Tchetgen (2018).
\newblock Identifying causal effects with proxy variables of an unmeasured
  confounder.
\newblock {\em Biometrika\/}~{\em 105\/}(4), 987--993.

\bibitem[\protect\citeauthoryear{Miller and Harrison}{Miller and
  Harrison}{2013}]{Mi2013}
Miller, J.~W. and M.~T. Harrison (2013).
\newblock A simple example of dirichlet process mixture inconsistency for the
  number of components.
\newblock {\em Advances in neural information processing systems\/}~{\em 26}.

\bibitem[\protect\citeauthoryear{M{\"u}ller, Quintana, Jara, and
  Hanson}{M{\"u}ller et~al.}{2015}]{Mu2015}
M{\"u}ller, P., F.~A. Quintana, A.~Jara, and T.~Hanson (2015).
\newblock {\em Bayesian nonparametric data analysis}, Volume~1.
\newblock Springer.

\bibitem[\protect\citeauthoryear{Nielsen, Gill, Andersen, and
  S{\o}rensen}{Nielsen et~al.}{1992}]{Ni1992}
Nielsen, G.~G., R.~D. Gill, P.~K. Andersen, and T.~I. S{\o}rensen (1992).
\newblock A counting process approach to maximum likelihood estimation in
  frailty models.
\newblock {\em Scandinavian journal of Statistics\/}~{\em 19\/}(1), 25--43.

\bibitem[\protect\citeauthoryear{Orihara, Fukuma, Ikenoue, and Taguri}{Orihara
  et~al.}{2024}]{Or2022}
Orihara, S., S.~Fukuma, T.~Ikenoue, and M.~Taguri (2024).
\newblock Likelihood-based instrumental variable methods for cox proportional
  hazards model.
\newblock {\em Japanese Journal of Statistics and Data Science\/}, 1--32.

\bibitem[\protect\citeauthoryear{Orihara and Goto}{Orihara and
  Goto}{2024}]{Or2024a}
Orihara, S. and A.~Goto (2024).
\newblock Comparison of instrumental variable methods with continuous exposure
  and binary outcome: A simulation study.
\newblock {\em Journal of Epidemiology\/}, JE20230271.

\bibitem[\protect\citeauthoryear{Orihara, Goto, and Taguri}{Orihara
  et~al.}{2023}]{Or2023}
Orihara, S., A.~Goto, and M.~Taguri (2023).
\newblock Instrumental variable estimation of causal effects with applying some
  model selection procedures under binary outcomes.
\newblock {\em Behaviormetrika\/}~{\em 50\/}(1), 241--262.

\bibitem[\protect\citeauthoryear{Shi, Miao, Nelson, and Tchetgen~Tchetgen}{Shi
  et~al.}{2020}]{Sh2020}
Shi, X., W.~Miao, J.~C. Nelson, and E.~J. Tchetgen~Tchetgen (2020).
\newblock Multiply robust causal inference with double-negative control
  adjustment for categorical unmeasured confounding.
\newblock {\em Journal of the Royal Statistical Society Series B: Statistical
  Methodology\/}~{\em 82\/}(2), 521--540.

\bibitem[\protect\citeauthoryear{Sinha, Ibrahim, and Chen}{Sinha
  et~al.}{2003}]{Si2003}
Sinha, D., J.~G. Ibrahim, and M.-H. Chen (2003).
\newblock A bayesian justification of cox's partial likelihood.
\newblock {\em Biometrika\/}~{\em 90\/}(3), 629--641.

\bibitem[\protect\citeauthoryear{Terza, Basu, and Rathouz}{Terza
  et~al.}{2008}]{Te2008}
Terza, J.~V., A.~Basu, and P.~J. Rathouz (2008).
\newblock Two-stage residual inclusion estimation: addressing endogeneity in
  health econometric modeling.
\newblock {\em Journal of health economics\/}~{\em 27\/}(3), 531--543.

\bibitem[\protect\citeauthoryear{Wade, Carslake, Sattar, Davey~Smith, and
  Timpson}{Wade et~al.}{2018}]{Wa2018}
Wade, K.~H., D.~Carslake, N.~Sattar, G.~Davey~Smith, and N.~J. Timpson (2018).
\newblock Bmi and mortality in uk biobank: revised estimates using mendelian
  randomization.
\newblock {\em Obesity\/}~{\em 26\/}(11), 1796--1806.

\bibitem[\protect\citeauthoryear{Wooldridge}{Wooldridge}{2014}]{Wo2014}
Wooldridge, J.~M. (2014).
\newblock Quasi-maximum likelihood estimation and testing for nonlinear models
  with endogenous explanatory variables.
\newblock {\em Journal of Econometrics\/}~{\em 182\/}(1), 226--234.

\end{thebibliography}

\newpage

\begin{landscape}

\begin{table}[H]
\begin{center}
\caption{Summary of simulation settings}
\label{tab0}
\begin{tabular}{ccc|c|l}\hline
{\bf Setting}&{\bf IV strength}&{\bf Confounder strength}&{\bf Scenario \#}&{\bf Parameter settings}\\\hline
``Easy to identify"&Strong&Strong&(a)&
\begin{tabular}{l}
$\alpha_{0}=16-8I\{U=1\}-14I\{U=2\}$\\
$\alpha_{z1}=1.5$\\
$\alpha_{uz2}=4-2I\{U=1\}-3I\{U=2\}$
\end{tabular}\\\cline{2-5}
&Weak&Strong&(b)&
\begin{tabular}{l}
$\alpha_{0}=16-8I\{U=1\}-14I\{U=2\}$\\
$\alpha_{z1}=0.5$\\
$\alpha_{uz2}=4-2I\{U=1\}-3I\{U=2\}$
\end{tabular}\\\cline{2-5}
&Strong&Weak&(c)&
\begin{tabular}{l}
$\alpha_{0}=16-8I\{U=1\}-14I\{U=2\}$\\
$\alpha_{z1}=1.5$\\
$\alpha_{uz2}=0.5(4-2I\{U=1\}-3I\{U=2\})$
\end{tabular}\\\cline{2-5}
&Weak&Weak&(d)&
\begin{tabular}{l}
$\alpha_{0}=16-8I\{U=1\}-14I\{U=2\}$\\
$\alpha_{z1}=0.5$\\
$\alpha_{uz2}=0.5(4-2I\{U=1\}-3I\{U=2\})$
\end{tabular}\\\hline
``Hard to identify"&Strong&Strong&(a)&
\begin{tabular}{l}
$\alpha_{0}=12-4I\{U=1\}-9I\{U=2\}$\\
$\alpha_{z1}=1.5$\\
$\alpha_{uz2}=4.5-2.5I\{U=1\}-3I\{U=2\}$
\end{tabular}\\\cline{2-5}
&Weak&Strong&(b)&
\begin{tabular}{l}
$\alpha_{0}=12-4I\{U=1\}-9I\{U=2\}$\\
$\alpha_{z1}=0.5$\\
$\alpha_{uz2}=4.5-2.5I\{U=1\}-3I\{U=2\}$
\end{tabular}\\\cline{2-5}
&Strong&Weak&(c)&
\begin{tabular}{l}
$\alpha_{0}=12-4I\{U=1\}-9I\{U=2\}$\\
$\alpha_{z1}=1.5$\\
$\alpha_{uz2}=0.5(4.5-2.5I\{U=1\}-3I\{U=2\})$
\end{tabular}\\\cline{2-5}
&Weak&Weak&(d)&
\begin{tabular}{l}
$\alpha_{0}=12-4I\{U=1\}-9I\{U=2\}$\\
$\alpha_{z1}=0.5$\\
$\alpha_{uz2}=0.5(4.5-2.5I\{U=1\}-3I\{U=2\})$
\end{tabular}\\\hline
\end{tabular}
\end{center}
\end{table}

\begin{table}[H]
\begin{center}
\caption{Summary of Estimates in ``Easy to identify" setting: The sample size and iteration time are $600$ and $200$, respectively. The true values of ``Other regression coefficients for the outcome model" and ``Variance parameter" are $0$, $0.5$, and $0.5$, respectively. In Scenarios (a) and (c), the true value of ``The regression coefficient for the exposure model" is $1.5$, whereas it is $0.5$ in Scenarios (b) and (d). Bias, empirical standard error (ESE), root mean squared error (RMSE), and coverage probability (CP) of the estimated log-hazard ratio in 200 iterations by estimation methods (``Method'' column) are summarized. For our proposed method, the posterior mean and standard deviation of other regression coefficients for the outcome model, the regression coefficient for the exposure model (i.e., regression coefficients for covariates), and the variance parameter are summarized.}
\label{tab1}
\scalebox{0.94}{$
\begin{tabular}{ll|cccc|cccc|cc|cc}\hline
{\bf Scenario}&{\bf Method}&\multicolumn{4}{|c}{\bf \begin{tabular}{c} Target parameter\\(log--hazard ratio)\end{tabular}}&\multicolumn{4}{|c}{\bf \begin{tabular}{c} Other regression\\coefficients for the\\outcome model\end{tabular}}&\multicolumn{2}{|c}{\bf \begin{tabular}{c} The regression\\coefficient for the\\exposure model\end{tabular}}&\multicolumn{2}{|c}{\bf \begin{tabular}{c} Variance\\parameter\end{tabular}}\\
&&{\bf Bias}&{\bf ESE}&{\bf RMSE}&{\bf CP}&{\bf Mean}&{\bf ESE}&{\bf Mean}&{\bf ESE}&{\bf Mean}&{\bf ESE}&{\bf Mean}&{\bf ESE}\\\hline
(a) & 1. Proposed & 0.004  & 0.034  & 0.034  & 0.985  & 0.085  & 0.108  & 0.498  & 0.095  & 1.567  & 0.050  & 0.554  & 0.165  \\
 & 2. Naive & 0.111  & 0.008  & 0.112  & 0.000  & - & - & - & - & - & - & - & - \\
 & 3. 2SLS & -0.014  & 0.266  & 0.267  & 0.855  & - & - & - & - & - & - & - & - \\
 & 4. 2SRI & 0.031  & 0.026  & 0.040  & 0.680  & - & - & - & - & - & - & - & - \\
 & 5. Infeasible & 0.000  & 0.040  & 0.040  & 0.955  & - & - & - & - & - & - & - & - \\\hline
(b) & 1. Proposed & 0.005  & 0.037  & 0.037  & 0.975  & 0.082  & 0.106  & 0.473  & 0.102  & 0.599  & 0.392  & 0.753  & 2.776  \\
 & 2. Naive & 0.116  & 0.007  & 0.116  & 0.000  & - & - & - & - & - & - & - & - \\
 & 3. 2SLS & -0.548  & 6.477  & 6.500  & 0.835  & - & - & - & - & - & - & - & - \\
 & 4. 2SRI & 0.029  & 0.028  & 0.041  & 0.650  & - & - & - & - & - & - & - & - \\
 & 5. Infeasible & 0.000  & 0.040  & 0.040  & 0.955  & - & - & - & - & - & - & - & - \\\hline
(c) & 1. Proposed & 0.004  & 0.045  & 0.045  & 0.985  & 0.092  & 0.078  & 0.539  & 0.137  & 1.568  & 0.047  & 0.606  & 0.224  \\
 & 2. Naive & 0.131  & 0.009  & 0.132  & 0.000  & - & - & - & - & - & - & - & - \\
 & 3. 2SLS & 0.008  & 0.057  & 0.057  & 0.890  & - & - & - & - & - & - & - & - \\
 & 4. 2SRI & 0.060  & 0.055  & 0.082  & 0.680  & - & - & - & - & - & - & - & - \\
 & 5. Infeasible & 0.001  & 0.051  & 0.051  & 0.945  & - & - & - & - & - & - & - & - \\\hline
(d) & 1. Proposed & 0.005  & 0.057  & 0.057  & 0.960  & 0.090  & 0.096  & 0.502  & 0.113  & 0.655  & 0.572  & 1.283  & 4.833  \\
 & 2. Naive & 0.137  & 0.009  & 0.137  & 0.000  & - & - & - & - & - & - & - & - \\
 & 3. 2SLS & 0.276  & 3.211  & 3.222  & 0.860  & - & - & - & - & - & - & - & - \\
 & 4. 2SRI & 0.058  & 0.056  & 0.081  & 0.640  & - & - & - & - & - & - & - & - \\
 & 5. Infeasible & 0.003  & 0.056  & 0.056  & 0.950  & - & - & - & - & - & - & - & - \\
\hline
\end{tabular}
$}
\end{center}
\end{table}

\begin{table}[H]
\begin{center}
\caption{Summary of Estimates in ``Easy to identify" setting: The sample size and iteration time are $1200$ and $200$, respectively. The true values of ``Other regression coefficients for the outcome model" and ``Variance parameter" are $0$, $0.5$, and $0.5$, respectively. In Scenarios (a) and (c), the true value of ``The regression coefficient for the exposure model" is $1.5$, whereas it is $0.5$ in Scenarios (b) and (d). Bias, empirical standard error (ESE), root mean squared error (RMSE), and coverage probability (CP) of the estimated log-hazard ratio in 200 iterations by estimation methods (``Method'' column) are summarized. For our proposed method, the posterior mean and standard deviation of other regression coefficients for the outcome model, the regression coefficient for the exposure model (i.e., regression coefficients for covariates), and the variance parameter are summarized.}
\label{tab2}
\scalebox{0.94}{$
\begin{tabular}{ll|cccc|cccc|cc|cc}\hline
{\bf Scenario}&{\bf Method}&\multicolumn{4}{|c}{\bf \begin{tabular}{c} Target parameter\\(log--hazard ratio)\end{tabular}}&\multicolumn{4}{|c}{\bf \begin{tabular}{c} Other regression\\coefficients for the\\outcome model\end{tabular}}&\multicolumn{2}{|c}{\bf \begin{tabular}{c} The regression\\coefficient for the\\exposure model\end{tabular}}&\multicolumn{2}{|c}{\bf \begin{tabular}{c} Variance\\parameter\end{tabular}}\\
&&{\bf Bias}&{\bf ESE}&{\bf RMSE}&{\bf CP}&{\bf Mean}&{\bf ESE}&{\bf Mean}&{\bf ESE}&{\bf Mean}&{\bf ESE}&{\bf Mean}&{\bf ESE}\\\hline
(a) & 1. Proposed & 0.004  & 0.026  & 0.027  & 0.975  & 0.084  & 0.086  & 0.478  & 0.121  & 1.563  & 0.034  & 0.490  & 0.083  \\
 & 2. Naive & 0.111  & 0.006  & 0.111  & 0.000  & - & - & - & - & - & - & - & - \\
 & 3. 2SLS & 0.007  & 0.042  & 0.043  & 0.875  & - & - & - & - & - & - & - & - \\
 & 4. 2SRI & 0.031  & 0.017  & 0.035  & 0.505  & - & - & - & - & - & - & - & - \\
 & 5. Infeasible & 0.001  & 0.029  & 0.029  & 0.945  & - & - & - & - & - & - & - & - \\\hline
(b) & 1. Proposed & 0.005  & 0.031  & 0.032  & 0.955  & 0.086  & 0.091  & 0.444  & 0.115  & 0.640  & 0.629  & 1.175  & 5.674  \\
 & 2. Naive & 0.116  & 0.005  & 0.116  & 0.000  & - & - & - & - & - & - & - & - \\
 & 3. 2SLS & 0.055  & 1.292  & 1.293  & 0.880  & - & - & - & - & - & - & - & - \\
 & 4. 2SRI & 0.032  & 0.016  & 0.036  & 0.445  & - & - & - & - & - & - & - & - \\
 & 5. Infeasible & 0.001  & 0.030  & 0.030  & 0.940  & - & - & - & - & - & - & - & - \\\hline
(c) & 1. Proposed & 0.002  & 0.043  & 0.043  & 0.945  & 0.092  & 0.070  & 0.500  & 0.113  & 1.668  & 0.716  & 1.344  & 5.772  \\
 & 2. Naive & 0.131  & 0.006  & 0.131  & 0.000  & - & - & - & - & - & - & - & - \\
 & 3. 2SLS & 0.007  & 0.040  & 0.041  & 0.875  & - & - & - & - & - & - & - & - \\
 & 4. 2SRI & 0.060  & 0.037  & 0.071  & 0.475  & - & - & - & - & - & - & - & - \\
 & 5. Infeasible & -0.002  & 0.039  & 0.039  & 0.950  & - & - & - & - & - & - & - & - \\\hline
(d) & 1. Proposed & -0.002  & 0.040  & 0.040  & 0.965  & 0.099  & 0.072  & 0.446  & 0.086  & 0.586  & 0.291  & 0.695  & 2.546  \\
 & 2. Naive & 0.136  & 0.006  & 0.137  & 0.000  & - & - & - & - & - & - & - & - \\
 & 3. 2SLS & 0.057  & 1.533  & 1.534  & 0.880  & - & - & - & - & - & - & - & - \\
 & 4. 2SRI & 0.059  & 0.036  & 0.070  & 0.500  & - & - & - & - & - & - & - & - \\
 & 5. Infeasible & -0.001  & 0.038  & 0.038  & 0.950  & - & - & - & - & - & - & - & - \\
\hline
\end{tabular}
$}
\end{center}
\end{table}

\begin{table}[H]
\begin{center}
\caption{Summary of Estimates in ``Hard to identify" setting: The sample size and iteration time are $600$ and $200$, respectively. The true values of ``Other regression coefficients for the outcome model" and ``Variance parameter" are $0$, $0.5$, and $0.5$, respectively. In Scenarios (a) and (c), the true value of ``The regression coefficient for the exposure model" is $1.5$, whereas it is $0.5$ in Scenarios (b) and (d). Bias, empirical standard error (ESE), root mean squared error (RMSE), and coverage probability (CP) of the estimated log-hazard ratio in 200 iterations by estimation methods (``Method'' column) are summarized. For our proposed method, the posterior mean and standard deviation of other regression coefficients for the outcome model, the regression coefficient for the exposure model (i.e., regression coefficients for covariates), and the variance parameter are summarized.}
\label{tab3}
\scalebox{0.94}{$
\begin{tabular}{ll|cccc|cccc|cc|cc}\hline
{\bf Scenario}&{\bf Method}&\multicolumn{4}{|c}{\bf \begin{tabular}{c} Target parameter\\(log--hazard ratio)\end{tabular}}&\multicolumn{4}{|c}{\bf \begin{tabular}{c} Other regression\\coefficients for the\\outcome model\end{tabular}}&\multicolumn{2}{|c}{\bf \begin{tabular}{c} The regression\\coefficient for the\\exposure model\end{tabular}}&\multicolumn{2}{|c}{\bf \begin{tabular}{c} Variance\\parameter\end{tabular}}\\
&&{\bf Bias}&{\bf ESE}&{\bf RMSE}&{\bf CP}&{\bf Mean}&{\bf ESE}&{\bf Mean}&{\bf ESE}&{\bf Mean}&{\bf ESE}&{\bf Mean}&{\bf ESE}\\\hline
(a) & 1. Proposed & 0.004  & 0.034  & 0.034  & 0.985  & 0.093  & 0.109  & 0.519  & 0.100  & 1.558  & 0.046  & 0.511  & 0.172  \\
 & 2. Naive & -0.026  & 0.012  & 0.029  & 0.445  & - & - & - & - & - & - & - & - \\
 & 3. 2SLS & 0.016  & 0.051  & 0.054  & 0.950  & - & - & - & - & - & - & - & - \\
 & 4. 2SRI & 0.029  & 0.023  & 0.037  & 0.765  & - & - & - & - & - & - & - & - \\
 & 5. Infeasible & -0.002  & 0.042  & 0.042  & 0.970  & - & - & - & - & - & - & - & - \\\hline
(b) & 1. Proposed & 0.003  & 0.036  & 0.036  & 0.995  & 0.093  & 0.114  & 0.489  & 0.137  & 0.555  & 0.053  & 0.535  & 0.288  \\
 & 2. Naive & -0.022  & 0.011  & 0.025  & 0.495  & - & - & - & - & - & - & - & - \\
 & 3. 2SLS & 4.689  & 66.751  & 66.915  & 0.970  & - & - & - & - & - & - & - & - \\
 & 4. 2SRI & 0.028  & 0.023  & 0.037  & 0.750  & - & - & - & - & - & - & - & - \\
 & 5. Infeasible & -0.001  & 0.041  & 0.041  & 0.950  & - & - & - & - & - & - & - & - \\\hline
(c) & 1. Proposed & 0.006  & 0.046  & 0.046  & 0.990  & 0.089  & 0.086  & 0.547  & 0.103  & 1.556  & 0.048  & 0.534  & 0.065  \\
 & 2. Naive & -0.025  & 0.013  & 0.028  & 0.535  & - & - & - & - & - & - & - & - \\
 & 3. 2SLS & 0.018  & 0.048  & 0.052  & 0.930  & - & - & - & - & - & - & - & - \\
 & 4. 2SRI & 0.061  & 0.047  & 0.077  & 0.720  & - & - & - & - & - & - & - & - \\
 & 5. Infeasible & 0.003  & 0.055  & 0.055  & 0.960  & - & - & - & - & - & - & - & - \\\hline
(d) & 1. Proposed & 0.002  & 0.050  & 0.050  & 0.985  & 0.095  & 0.092  & 0.501  & 0.104  & 0.557  & 0.049  & 0.569  & 0.318  \\
 & 2. Naive & -0.020  & 0.012  & 0.024  & 0.665  & - & - & - & - & - & - & - & - \\
 & 3. 2SLS & -0.101  & 1.727  & 1.730  & 0.960  & - & - & - & - & - & - & - & - \\
 & 4. 2SRI & 0.061  & 0.048  & 0.077  & 0.690  & - & - & - & - & - & - & - & - \\
 & 5. Infeasible & 0.001  & 0.055  & 0.055  & 0.945  & - & - & - & - & - & - & - & - \\
\hline
\end{tabular}
$}
\end{center}
\end{table}

\begin{table}[H]
\begin{center}
\caption{Summary of Estimates in ``Hard to identify" setting: The sample size and iteration time are $1200$ and $200$, respectively. The true values of ``Other regression coefficients for the outcome model" and ``Variance parameter" are $0$, $0.5$, and $0.5$, respectively. In Scenarios (a) and (c), the true value of ``The regression coefficient for the exposure model" is $1.5$, whereas it is $0.5$ in Scenarios (b) and (d). Bias, empirical standard error (ESE), root mean squared error (RMSE), and coverage probability (CP) of the estimated log-hazard ratio in 200 iterations by estimation methods (``Method'' column) are summarized. For our proposed method, the posterior mean and standard deviation of other regression coefficients for the outcome model, the regression coefficient for the exposure model (i.e., regression coefficients for covariates), and the variance parameter are summarized.}
\label{tab4}
\scalebox{0.94}{$
\begin{tabular}{ll|cccc|cccc|cc|cc}\hline
{\bf Scenario}&{\bf Method}&\multicolumn{4}{|c}{\bf \begin{tabular}{c} Target parameter\\(log--hazard ratio)\end{tabular}}&\multicolumn{4}{|c}{\bf \begin{tabular}{c} Other regression\\coefficients for the\\outcome model\end{tabular}}&\multicolumn{2}{|c}{\bf \begin{tabular}{c} The regression\\coefficient for the\\exposure model\end{tabular}}&\multicolumn{2}{|c}{\bf \begin{tabular}{c} Variance\\parameter\end{tabular}}\\
&&{\bf Bias}&{\bf ESE}&{\bf RMSE}&{\bf CP}&{\bf Mean}&{\bf ESE}&{\bf Mean}&{\bf ESE}&{\bf Mean}&{\bf ESE}&{\bf Mean}&{\bf ESE}\\\hline
(a) & 1. Proposed & 0.004  & 0.026  & 0.026  & 1.000  & 0.091  & 0.087  & 0.474  & 0.108  & 1.543  & 0.032  & 0.492  & 0.182  \\
 & 2. Naive & -0.027  & 0.009  & 0.028  & 0.105  & - & - & - & - & - & - & - & - \\
 & 3. 2SLS & 0.016  & 0.034  & 0.037  & 0.935  & - & - & - & - & - & - & - & - \\
 & 4. 2SRI & 0.027  & 0.015  & 0.032  & 0.675  & - & - & - & - & - & - & - & - \\
 & 5. Infeasible & 0.001  & 0.030  & 0.030  & 0.965  & - & - & - & - & - & - & - & - \\\hline
(b) & 1. Proposed & 0.001  & 0.027  & 0.027  & 0.980  & 0.100  & 0.086  & 0.442  & 0.114  & 0.543  & 0.036  & 0.492  & 0.275  \\
 & 2. Naive & -0.022  & 0.008  & 0.024  & 0.220  & - & - & - & - & - & - & - & - \\
 & 3. 2SLS & 0.034  & 0.250  & 0.253  & 0.935  & - & - & - & - & - & - & - & - \\
 & 4. 2SRI & 0.029  & 0.016  & 0.033  & 0.525  & - & - & - & - & - & - & - & - \\
 & 5. Infeasible & 0.000  & 0.030  & 0.030  & 0.970  & - & - & - & - & - & - & - & - \\\hline
(c) & 1. Proposed & -0.005  & 0.037  & 0.037  & 0.975  & 0.104  & 0.073  & 0.494  & 0.083  & 1.547  & 0.035  & 0.529  & 0.172  \\
 & 2. Naive & -0.026  & 0.009  & 0.027  & 0.165  & - & - & - & - & - & - & - & - \\
 & 3. 2SLS & 0.015  & 0.033  & 0.036  & 0.905  & - & - & - & - & - & - & - & - \\
 & 4. 2SRI & 0.058  & 0.032  & 0.066  & 0.540  & - & - & - & - & - & - & - & - \\
 & 5. Infeasible & -0.001  & 0.040  & 0.040  & 0.970  & - & - & - & - & - & - & - & - \\\hline
(d) & 1. Proposed & -0.005  & 0.037  & 0.037  & 0.975  & 0.105  & 0.072  & 0.454  & 0.080  & 0.546  & 0.031  & 0.514  & 0.212  \\
 & 2. Naive & -0.021  & 0.009  & 0.023  & 0.345  & - & - & - & - & - & - & - & - \\
 & 3. 2SLS & 0.005  & 0.133  & 0.133  & 0.935  & - & - & - & - & - & - & - & - \\
 & 4. 2SRI & 0.058  & 0.030  & 0.066  & 0.500  & - & - & - & - & - & - & - & - \\
 & 5. Infeasible & -0.004  & 0.039  & 0.039  & 0.960  & - & - & - & - & - & - & - & - \\
\hline
\end{tabular}
$}
\end{center}
\end{table}

\end{landscape}

\begin{table}[H]
\begin{center}
\caption{Summary of constructed clusters in ``Hard to identify" setting and Scenario (a).}
\label{tab5}

\begin{tabular}{c|ccc}\hline
&\multicolumn{3}{|c}{\bf True cluster ID}\\
{\bf Estimated cluster ID}&$U=0$&$U=1$&$U=2$\\\hline
1 & 0 & 24 & 0 \\
2 & 66 & 0 & 0 \\
3 & 60 & 0 & 0 \\
4 & 26 & 5 & 0 \\
5 & 0 & 19 & 0 \\
6 & 49 & 0 & 0 \\
7 & 0 & 28 & 0 \\
8 & 51 & 0 & 0 \\
9 & 58 & 0 & 0 \\
10 & 51 & 0 & 0 \\
11 & 57 & 0 & 0 \\
12 & 53 & 0 & 0 \\
13 & 49 & 0 & 0 \\
14 & 0 & 23 & 0 \\
15 & 52 & 1 & 0 \\
16 & 54 & 0 & 0 \\
17 & 0 & 27 & 0 \\
18 & 0 & 36 & 0 \\
19 & 0 & 0 & 26 \\
20 & 0 & 21 & 0 \\
21 & 0 & 25 & 0 \\
22 & 0 & 27 & 0 \\
23 & 0 & 21 & 0 \\
24 & 0 & 28 & 0 \\
25 & 0 & 21 & 0 \\
26 & 0 & 15 & 0 \\
27 & 0 & 0 & 14 \\
28 & 0 & 21 & 0 \\
29 & 0 & 0 & 22 \\
30 & 0 & 24 & 0 \\
31 & 0 & 0 & 21 \\
32 & 0 & 0 & 15 \\
33 & 0 & 0 & 9 \\
34 & 0 & 0 & 17 \\
35 & 0 & 0 & 14 \\
36 & 0 & 0 & 20 \\
37 & 0 & 21 & 0 \\
38 & 0 & 0 & 19 \\
39 & 0 & 0 & 10 \\\hline
\end{tabular}
\end{center}
\end{table}

\newpage


\begin{landscape}
\begin{figure}[H]
\begin{center}
\begin{tabular}{c}
\includegraphics[width=22cm]{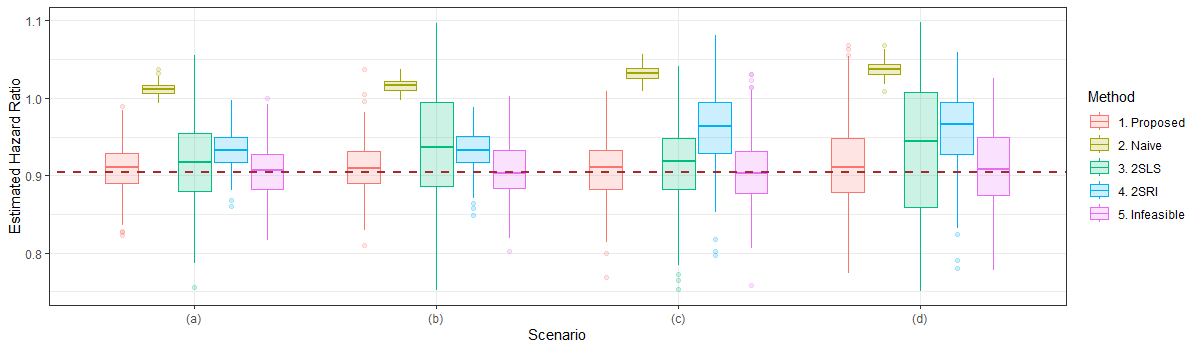}\\
\includegraphics[width=22cm]{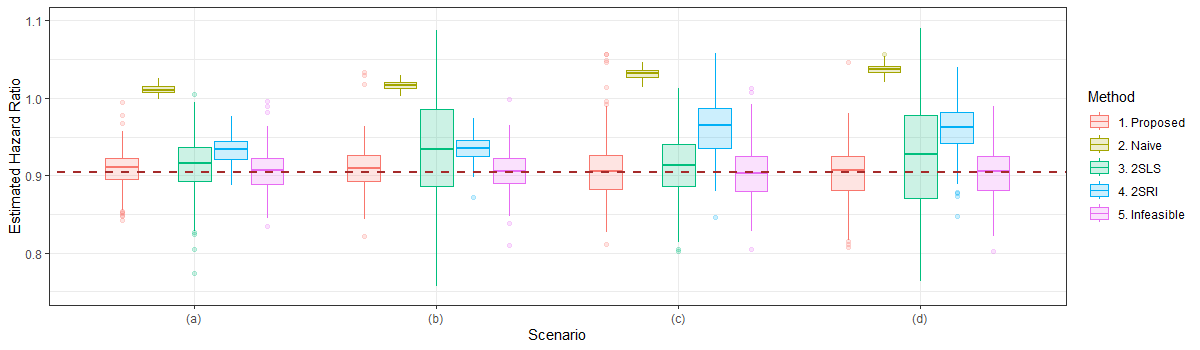}
\end{tabular}
\caption{Box plots of hazard ratio estimates for each method in ``Easy to identify" setting: The iteration time is $200$. The true values of hazard ratio is $\exp\{-0.1\}\approx0.905$.}
\label{fig6}
\end{center}
{\footnotesize
Upper figure: the sample is $600$; lower figure: the sample is $1200$.
}
\end{figure}

\begin{figure}[H]
\begin{center}
\begin{tabular}{c}
\includegraphics[width=22cm]{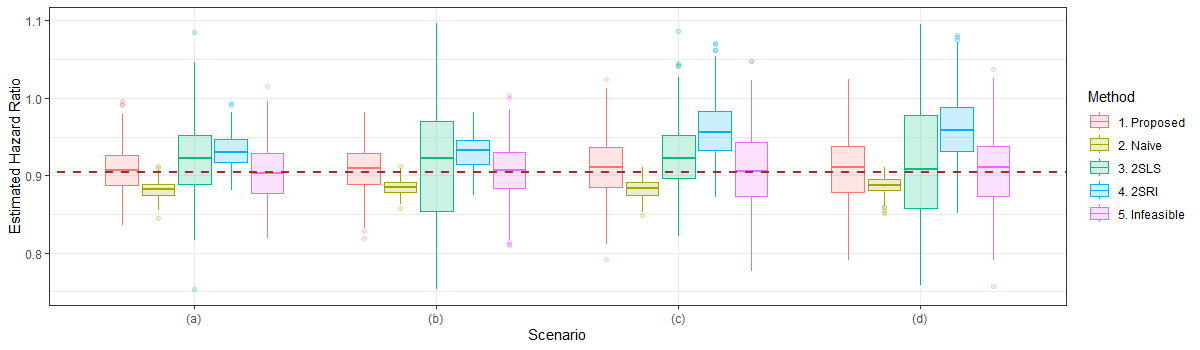}\\
\includegraphics[width=22cm]{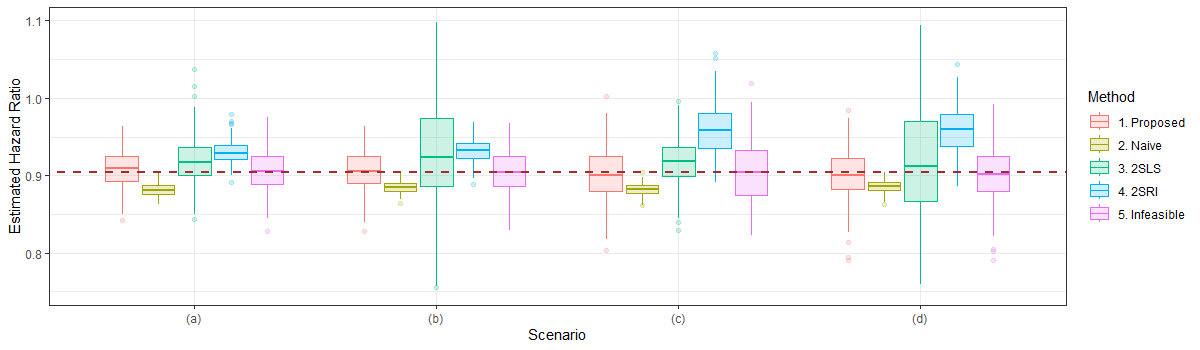}
\end{tabular}
\caption{Box plots of hazard ratio estimates for each method in ``Hard to identify" setting: The iteration time is $200$. The true values of hazard ratio is $\exp\{-0.1\}\approx0.905$.}
\label{fig7}
\end{center}
{\footnotesize
Upper figure: the sample is $600$; lower figure: the sample is $1200$.
}
\end{figure}

\begin{figure}[H]
\begin{center}
\begin{tabular}{c}
\includegraphics[width=22cm]{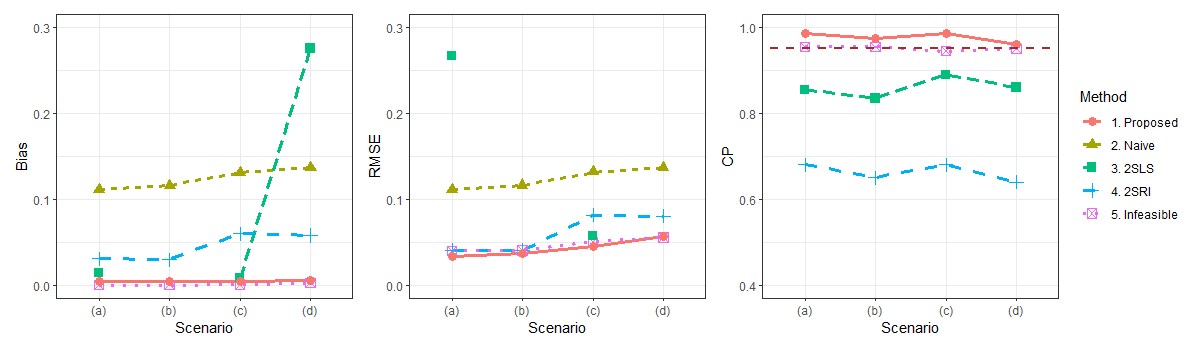}\\
\includegraphics[width=22cm]{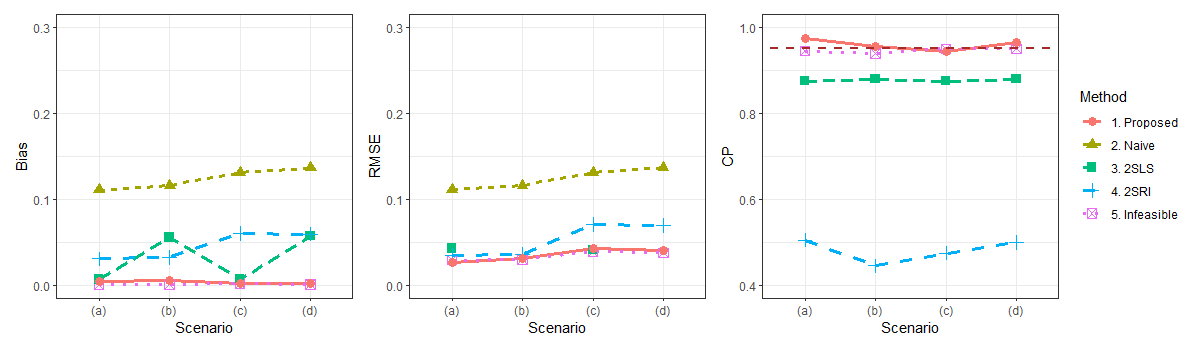}
\end{tabular}
\caption{Some plots of statistics for each method in ``Easy to identify" setting: The iteration time is $200$. Bias, root mean squared error (RMSE), and coverage probability (CP) of the estimated hazard ratio in 200 iterations by estimation methods are summarized.}
\label{fig8}
\end{center}
{\footnotesize
Upper figure: the sample is $600$; lower figure: the sample is $1200$.\\
Extreme values are not plotted: some Bias and RMSE of 2SLS, and CP of Naive estimator.
}
\end{figure}

\begin{figure}[H]
\begin{center}
\begin{tabular}{c}
\includegraphics[width=22cm]{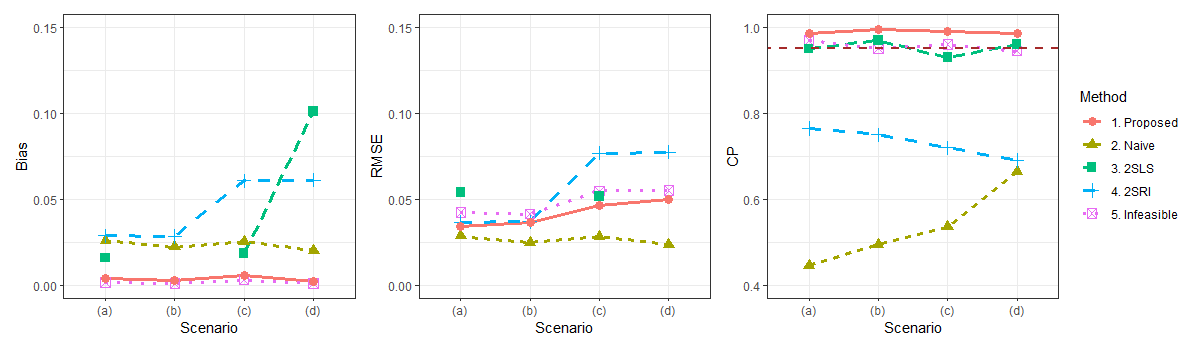}\\
\includegraphics[width=22cm]{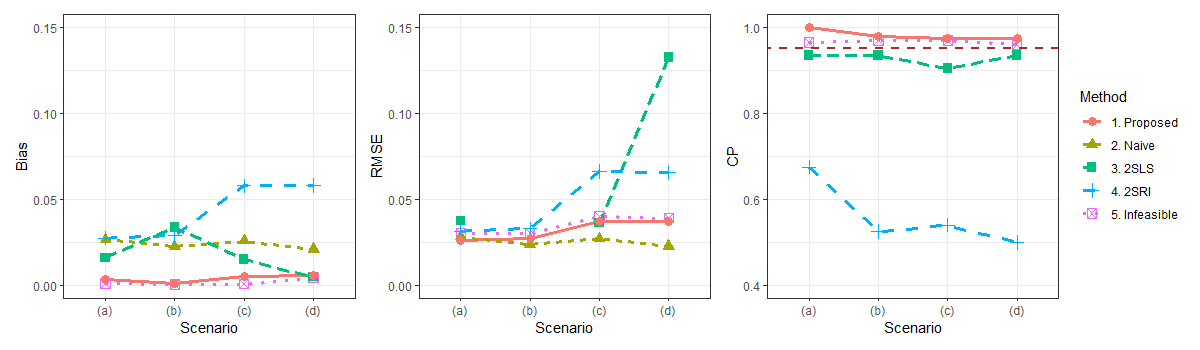}
\end{tabular}
\caption{Some plots of statistics for each method in ``Hard to identify" setting: The iteration time is $200$. Bias, root mean squared error (RMSE), and coverage probability (CP) of the estimated hazard ratio in 200 iterations by estimation methods are summarized.}
\label{fig9}
\end{center}
{\footnotesize
Upper figure: the sample is $600$; lower figure: the sample is $1200$.\\
Extreme values are not plotted: some Bias and RMSE of 2SLS, and CP of Naive estimator.
}
\end{figure}

\end{landscape}

\begin{figure}[H]
\begin{center}
\begin{tabular}{c}
\includegraphics[width=16cm]{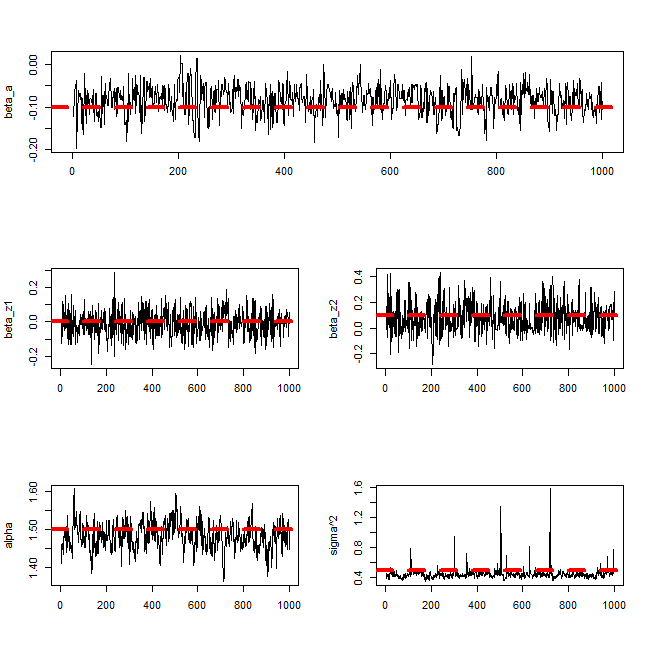}
\end{tabular}
\caption{Sampling plot for proposed method under ``Hard to identify" setting and Scenario (a) in a dataset}
\label{fig10}
\end{center}
\end{figure}

\newpage
\appendix
\section{Detailed Sampling Algorithms for $S$ and $\gamma$}
\label{appa}
Using the discussion, for example in \cite{Da2017}, 
the full conditional distribution of $s_i$ is 
$$
P(s_i=k|\cdot)\propto p(S^{i\to k};\gamma)\ell_{ik}(\beta_{a}, \bld{\beta}_{x})\phi\Big(A_i; \alpha_{0k}+\bld{v}_i^{\top}\bld{\alpha}_{vk}+\bld{z}_i^{\top}\bld{\alpha}_{z}, \sigma_k^2\Big), \ \ \  k=1,\ldots,K_n+1,
$$
where $S^{i\to k}$ denotes a sequence $s_1,\ldots,s_n$ with $s_i=k$ and the other variables set to the current assignment, and  
$p(S^{i\to k};\gamma)$ is the joint probability obtained by the Chinese restaurant representation  (\ref{eq:DP2}).
For $k=K_n+1$, the parameter $(\alpha_{0k},\bld{\alpha}_{vk},\sigma_{k}^2)$ are generated from their prior distribution at each MCMC iteration. In short, clustering is achieved either by creating a new cluster for subject $i$, or by assigning the subject to an existing cluster with some probabilities. By iterating the sampling from the posterior, clusters $S$ are automatically created, adjusted by the hyper-parameter $\gamma$. In the simulation experiments, we assume $G_{0}$ follows a normal distribution to simplify the calculation. 

In the simulation experiments, the precision parameter $\gamma$ is sampled using the same strategy as \cite{Es1998}. Specifically, it is assumed that $\gamma$ has a Gamma prior distribution: $\gamma\sim Gamma(a,b)$. Additionally, we introduce a latent variable (parameter) $\eta$. Under this setting, the posteriors of these variables are described as follows:
\begin{description}
\item[Posterior for $\gamma$]
$$
\gamma|\eta,I^{*}\sim \pi Gamma(a+I^{*},b-\log(\eta))+(1-\pi)Gamma(a+I^{*}-1,b-\log(\eta)),
$$
where $\pi=e^{\lambda}/(1+e^{\lambda})$, $\lambda=(a+I^{*}-1)/I^{*}\{b-\log(\eta)\}$,
and $I^{*}$ is the current cluster size.
\item[Posterior for $\eta$]
$$
\eta|\gamma,I^{*}\sim {\rm Beta}(a+1,n).
$$
\end{description}
In the simulation experiments, $a$ and $b$ are set as noted in the footnote of each table.

\section{Overcoming Homogeneous Treatment Effect Assumption and its Limitations}
\label{appb}
By modifying the model (\ref{for1_2}), we can consider a more complicated model that the parameter $\bld{\beta}$ may be different in each cluster:
\begin{align}
\lambda_{i}(t)=\lambda_{0i}(t)\exp\left(\tilde{\bld{x}}_{i}^{\top}\bld{\beta}_{i}\right). \label{for1_2_2}
\end{align}
For this model, the posterior sampling algorithm described in Section 3.2 needs to be slightly modified, however, sampling for $\bld{\beta}_{i}$ can be considered in the same manner. In (\ref{for1_2_2}), the treatment effect $\beta_{ai}$ may change for each (unknown) cluster, i.e., unmeasured confounders. In other words, (\ref{for1_2_2}) potentially allows for treatment effect heterogeneity with respect to unmeasured confounders. This is an important difference compared to ordinary IV methods since they commonly need homogeneous treatment assumption, except for methods under monotonicity assumption \citep{An1996}.

However, the interpretation of the model (\ref{for1_2_2}) is somewhat difficult since the clustering algorithm is completely nonparametric. When there are some measured covariates related to the unmeasured confounders, the clusters may be interpreted by confirming the measured covariates. For instance, a cluster is consist of many male and over 65 years old, whereas the other cluster is consist of the other subjects. Also, due to the noncollapsibility of the CPHM, the integrated log hazard ratio for each cluster cannot be interpreted simply. Summarizing the above discussions, we recommend modeling the CPHM as (\ref{for1_2}) for better interpretability of analysis results.

\end{document}